# Explaining herding and volatility in the cyclical price dynamics of urban housing markets using a large scale agent-based model

Kirill S. Glavatskiy[1]*, Mikhail Prokopenko[1], Adrian Carro[2], Paul Ormerod[3], Michael Harré[1]†.

[1] Centre for Complex Systems, The University of Sydney, Australia.
[2] Institute for New Economic Thinking at the Oxford Martin School, University of Oxford, UK, and Banco de España, Spain.
[3] Department of Computer Science, University College London, and Algorithmic Economics Ltd.
* e-mail for correspondence: k.s.glavatskiy@gmail.com
† e-mail for correspondence: michael.harre@sydney.edu.au.

**Abstract**

Urban housing markets, along with markets of other assets, universally exhibit periods of strong price increases followed by sharp corrections. The mechanisms generating such non-linearities are not yet well understood. We develop an agent-based model populated by a large number of heterogeneous households. The agents' behavior is compatible with economic rationality, with the trend-following behavior found to be essential in replicating market dynamics. The model is calibrated using several large and distributed datasets of the Greater Sydney region (demographic, economic and financial) across three specific and diverse periods since 2006. The model is not only capable of explaining price dynamics during these periods, but also reproduces the novel behavior actually observed immediately prior to the market peak in 2017, namely a sharp increase in the variability of prices. This novel behavior is related to a combination of trend-following aptitude of the household agents (rational herding) and their propensity to borrow.

**Summary**

Ruptured dynamics in housing markets results from the combination of rational herding and an unrestrained capacity to borrow.

**Introduction**

Urban housing markets in developed economies around the world exhibit a key characteristic in common. There are periods when house prices rise very rapidly, which are followed by sharp falls. In common parlance, this is often referred to as the "boom-bust" cycle. We illustrate this below in Fig. 1 which shows timeseries price index data from 2000 for selected cities worldwide.

The motivation of this paper is to develop a model based upon the interactions of heterogeneous agents which, as a fundamental feature, is able to generate these types of fluctuations in housing markets. We illustrate this general feature with specific simulations of the model calibrated on the data for the housing market in Greater Sydney. We choose Sydney because over the past 20 years – the period over which detailed and high resolution housing market and microeconomic data become available – the market has experienced not one but two periods of notable price corrections. In addition, the very detailed data available for Sydney identifies a very marked increase in the variability of monthly house prices in the period immediately prior to the most recent downturn at the end of the 2010s. This is a further key feature that an adequate model of the housing market ought to be able to generate.

These boom-bust cycles have been considered in general (*1*) but little consideration has been given to the feedbacks which generate the sensitivity of the cyclical behavior with respect to both exogenous and endogenous factors. We argue in this paper that trend-following market sentiment is a prominent feature of such feedback dynamics, often leading to non-linear amplification of price behavior.

Agent based models (ABM) are one of the most significant developments to emerge in economic modelling in recent years and they have been proposed as an alternative solution to the modelling of complex economic dynamics that formal or aggregate level models are not well suited for (*2-4*). Their strength lies in the micro-interactions between individual decision-making agents (*5*) each with their own characteristics, for example businesses, households, or consumers, as well as spatial characteristics that geographically localize these interactions (*6, 7*), which are affected by banks' fiscal policies (*8-10*) and macroeconomics conditions (*11-13*). These agent-to-agent interactions show complex (*14*), non-linear effects such as tipping-points (*15, 16*), boom-bust cycles (*17*), and chaos (*18, 19*), as well as the equilibrium dynamics predicted by classical models. More recently, an ABM was shown to outperform predictions made by several benchmark models which were based on vector auto-regression and dynamic stochastic general equilibrium approaches (*20*). Such approaches are prominent in mainstream economics, commonly used by central banks for policy decision-making, providing motivation for the use of ABMs to complement current economic modelling and forecasting techniques.

A key advantage is that, unlike formal approaches that often rely on an assumption of equilibrium (*21*), the results of ABM simulations can be either in or out of equilibrium. Importantly, the results of simulations can be interrogated in order to understand how the interactions of heterogeneous agents at the micro-level drive and give rise to the macro-level system dynamics. This provides researchers with vital micro-level insights into the causes of macro-level outcomes such as volatility clustering, fat tails, income distributions, and autocorrelations. These insights can then be checked against further micro-level data in order to be verified. As such ABMs have a much wider range of applicability than traditional approaches.

This breadth of applicability comes at a cost though; the necessary micro-economic data, over relatively short time periods, is often hard to obtain, the quantity of data is often difficult to manage, and the quality of the data itself can be heterogeneous. At the same time the decision-making algorithms of the agents need to be well founded in micro-economic principles, as do the interactions between the agents that influence these decisions. While these factors make the task difficult, significant advances have been made recently in the collection, curation, and deployment of government data, as well as in the commercial availability of data collected by private industries. For example many non-economic ABMs need to integrate disparate data sets and yet have had significant success in simulating complex social dynamics such as the collapse of societies (*22*) and pandemic spread and intervention strategies across a country or the globe (*23, 24*).

In housing economics there have also been recent successes but there are still only a few models that have used high-resolution household level data to successfully model large sectors of the market. The ABM developed by Axtell *et al* (*17, 25, 26*) for Washington DC replicated the features of the rise and fall (bubble and crash) dynamics seen in the house price index for Washington DC during the 1997-2009 period, including the period of the global financial crisis. The subsequent ABMs investigated various aspects of housing markets, such as the long-term macroeconomic aspects of a repeated rise and fall dynamic (*27, 28*), households' creditworthiness conditions (*29*), income segregation (*30*), effects of loans and mortgage securitization (*31, 32*) and general financial instability (*33*).

In this paper we use an ABM approach to model the Greater Sydney region at the individual 'household accounts' level. In this model each household's budget is individually represented: their income, tax, discretionary spending, housing expenses as well as a range of housing market and macroeconomic factors (see Materials and Methods). The transition of a household from renting to home ownership to owning an investment property, or in the other direction divesting to owning one home or renting, is modelled heterogeneously based on the individual household's ability to afford the mortgage costs and the willingness of the bank to provide a mortgage for the purchase.

We focus on three specific periods in the Australian housing market. First, the period 2006-2009 which contains the global financial crisis. Although Australia was one of the few Western economies not to experience a serious general economic recession, as Fig. 2 shows there was a fairly marked but short-lived market downturn. We also examine the period 2011-2014 when the housing market was recovering very slowly, in order to show that our model is capable of explaining a range of features of the housing market.

We calibrate our model to these two periods. With one important exception, we use the parameters calibrated on these two periods and apply them to the third period, 2016-2019, which exhibits the most recent price correction preceded by very substantial market volatility. The only model parameter that was changed is the trend-following aptitude, attributed to household agents. This parameter is shown to affect the market volatility in the context of current market and economic conditions which include lower interest rates and higher market confidence (propensity to borrow) quantified via heterogeneous mortgage-income distributions.

Our model is able to track the actual aggregate house price index based on heterogeneous budgetary constraints and buying and selling decisions of individual households. In addition, the model captures key qualitative dynamics such as market turning points and market volatility.

**Results**

We present the results of an agent-based model which simulates the price dynamics of the Greater Sydney housing market. The model is calibrated by simulating two periods 2006-2009 and 2011-2014, corresponding to the Censuses held in Australia in 2006 and 2011. We use the resulting parameters, as described above, to simulate the market from July 2016 to December 2018, i.e. over a period of 30 months. The starting date of the simulation is aligned with the Australian Census held in 2016, which ensures the best representation of the real population.

The ABM simulates the decision making dynamics of a large number of household agents (200 thousand). This number is necessary in order to capture fully the heterogeneous nature across major variables of the real population structure of Greater Sydney (2 million households) taken from Australian Census.

One fundamental feature representing heterogeneity of agents is quantification of the relationship between the income of the households and their propensity to borrow. Particularly, at the high end of the income distribution, this propensity varies substantially. We show that this relationship, when coupled with the trend-following mechanism, is essential in order to capture the large increase in variability of the market seen immediately prior to the turning point in 2018.

The agents implement their decisions, acting in discrete time steps according to a certain algorithm. In particular, they interact with each other, following the same budget-balancing and investment rules, which model typical decisions of real households. Every agent possesses certain attributes (e.g. income, liquid wealth, residence), which affect their decisions and, in turn, are affected by these decisions. Furthermore, the agents interact with the environment (e.g. the bank) subject to certain constraints, such as mortgage rate and financial prudence policies. Each simulation run results in a multivariate time series of key market indicators, such as price index, proportion of investing households, foreign participation, etc. The output of the model is an aggregate result of the decisions of the interacting heterogeneous household agents.

As an input the model uses a distinct set of the *external* environmental parameters, the values of which represent the state of the real market at the beginning of the simulation period. They include the external financial constraints (Table 1) and statistical structure of the population (Table 2). The temporal dependence and the statistical distributions are shown in Figures S1-S5 of Supplementary Materials. Furthermore, the model is characterized by a set of *internal* algorithmic parameters (Table 3), which do not vary in time. To reiterate, their values were calibrated for the periods of 2006-2009 and 2011-2014 and used for the period of 2016-2019. The internal parameters are thus the same for all three periods and are used by the algorithm as described in Materials and Methods. Finally, each of the periods is characterized by a single *feedback* parameter which couples the external and internal aspects of the simulated market. It reflects the potential of rational agents to follow so-called "herding" behavior (*34*) and quantifies their desire to follow the price trend; it will be referred to as the *trend-following aptitude*. The value of the trend-following aptitude is calibrated to the observable price dynamics and is differentiated across historical periods.

*Price trajectory*

The results for the periods of 2006-2009, 2011-2014, and 2016-2019 are shown in Fig. 3. Each simulated period is represented by ensemble trajectories (left) and the histogram (right) of the yearly moving average price. The corresponding values of the monthly average deal price and yearly moving average deal price from the *SIRCA / CoreLogic* data are shown for reference as well.

We see that the simulated price trajectories in the period of 2006-2009 exhibit a strong cyclical dynamic, showing stable increase followed by rapid change of the trend and stable decrease. The magnitude and the width of the cycle repeat those observed in reality, with the simulated price trajectory being slightly ahead of the actual one. All simulated trajectories follow a similar rising-then-declining trend. The price volatility within a single trajectory as well as variability between different trajectories is small. Still, each trajectory represents a separate market dynamic. This can be verified by absence of correlations between the price at the beginning of the simulated period and the price at the end of the simulated period, as illustrated in Fig. S6 (left). For this particular period the simulated trajectories follow a similar profile within a narrow band.

The simulated price trajectories in the period of 2011-2014 exhibit low but steady growth. This is also similar to the actual price dynamics, except for a short dip during 2012. The ensemble variability is again low, while individual trajectories show little correlation with significant heterogeneity between the start and the end of the period (Fig. S6, middle).

*Price variability*

In contrast to the two previous periods, the simulated price trajectories in the period of 2016-2019 exhibit high ensemble variability. This indicates that the market conditions facilitate a broad range of possible trajectory realizations and there is a high degree of uncertainty in the price dynamics. Yet, the ensemble average of the trajectories reproduces the actual dynamics well. In particular, it shows significant growth, which eventually plateaus in mid-2017 and is followed by a slow decline until 2019. Furthermore, the rising-then-declining dynamics in the period of 2006-2009 and of 2016-2019 are similar, while the latter one shows much larger variability.

We have also found that there are two parameters, which affect most the price variability: the mortgage rate and the heterogeneous relationship between mortgages and income, which reflects the propensity to borrow. This effect is only observed in combination with a high trend-following aptitude. We examined a range of possible factors potentially contributing to variability and verified that changing the majority of financial conditions (which are accounted by "external" parameters) from the 2016-2019 values to 2011-2014 values (when the variability was low) does not reduce the ensemble variability of the 2016-2019 price. Importantly, changing the trend-following aptitude alone does not reduce the ensemble variability either. However, changing some of the financial conditions *together* with reducing the trend-following aptitude does reduce the ensemble variability. We particularly note that we have found that the high ensemble variability is caused by one of the two combinations (see Fig. S7, bottom): i) low mortgage rate (the 2016-2019 rates are lower than the 2011-2104 rates) *and* high trend-following aptitude; ii) the high propensity to borrow *and* high trend-following aptitude. Altering other parameters, such as population level, housing stock, initial price magnitude, distribution of wealth, income, mortgage, or overseas investment activity does not affect much the price variability within the ensemble (although it does change the price trend).

*Trend-following aptitude*

In addition to the environmental parameters the model uses a feedback parameter which reflects the agent's desire to follow the price trend. This affects the agent's bid price when they participate in the market. The price trend is represented by the annual change of the house price index, $\Delta_{HPI}$, which is positive if the price goes up, and negative otherwise, while the price index itself is calculated by the BMN methodology (*35*). The price trend contributes to the bid price decision, which imposes a feedback in the model and couples the fixed internal decision model with the dynamic external conditions. The bid price $P_b$ is modelled as $P_b = A/(B - h\,\Delta_{HPI})$, where $h$ is the trend-following aptitude, while $A$ and $B$ are functions of variables for which the exact expression is given by Eq. (3) of Materials and Methods. The parameter $h$ shows how sensitive the agents' desire to buy a house is to the overall price trend. If $h = 0$ then the agents are completely indifferent to the observable price dynamics, and the magnitude of $h$ quantifies the strength of the feedback.

The value of the trend-following aptitude is the same for all agents. One may draw an analogy with thermodynamics in line with social physics research (*36*). For example, the trend-following aptitude may be seen as a collective state variable of the market. A higher magnitude of the aptitude would correspond to a more "heated" market, while a lower magnitude of the aptitude would correspond to a more "cooled" market.

One of the purposes of this work is to identify how the trend-following aptitude may be related to the actual market dynamics. To do this we assume that the value of the aptitude is fixed during a single simulation period. Yet, we allow the aptitude to be different for different simulated periods. The actual value of the trend-following aptitude is calibrated to the actual price dynamics, minimizing the distance between the actual and simulated prices over each period.

The resulting values for each period are the following: $h_{2006} = 0.45$ for the period of 2006-2009, $h_{2011} = -0.10$ for the period of 2011-2014, and $h_{2016} = 0.65$ for the period of 2016-2019.

The aptitude value is quite high for the period of 2006-2009, which means that the simulated market is "heated" and agents tend to follow the collective behavior more that the individual one. This agrees with the perception of the actual market at the time.

The magnitude of the aptitude is the lowest for the period of 2011-2014, having negative sign and moderate magnitude. This means that the simulated market is "cooled" and agents are not expecting the price to increase. Rather, they tend to ignore the price trend or even expect the price to reverse. The Sydney market was transitioning from a peak in 2011 to a dip in 2012, followed by a steady growth in the subsequent five years. A negative value of the obtained aptitude indicates the agents' caution and skepticism towards trend-following. However, the model still reproduces the modest growth which was seen in the actual market. The other features of the model, reflecting fundamental rather than speculative aspects of the market, are sufficient to override the negative aptitude. This suggests that increasing price on the housing market in 2011-2014 was driven by financial factors rather than people's collective behavior.

For the period of 2016-2019 the value of the aptitude is the highest, which indicates that the market is "super-heated". This value, in combination with low interest rate and/or high propensity to borrow, is, as we explained above, the mechanism which generates the large ensemble variability of the market in the simulations, with an abundance of outlier trajectories. This reflects the large variability in the actual market.

# Discussion

In this work we have developed an agent-based model which simulates universal features of urban housing markets observed across the world.

Our empirical realization of the model is based on demographic data and market conditions for Greater Sydney over three different historical periods taken from the last 15 years. Using a deterministic algorithm for interactions between ca. 200 000 heterogeneous agents we reproduce the actual market dynamics for these three periods. The first period coincides with the global financial crisis and a corresponding price increase and correction in the Sydney housing market. The second period covers a completely different historical experience in which prices initially fell slightly and then recovered moderately.

The third period covers the end of a period of rapid growth in the Sydney market followed by its subsequent decline. Crucially, the variability of the actual prices increased very markedly immediately prior to the turning point and the model replicates this behavior. We note that, with the exception of the trend-following parameter, the parameters used in the simulations of the third period are those obtained from the calibration using the first two periods. In other words, the model is capable of generating novel behavior not observed in the real world during the calibration periods. This novel behavior does concur with the actual market dynamics in the third period, given initial market and economic conditions.

The households in the model follow rules of behaviour which are consistent with the rational postulates of economics. Even when they exhibit herding behaviour in following market sentiment, this is a phenomenon which is fully compatible with individual rationality. The key insights of the model are obtained by combining such postulates of behaviour with data-driven approaches. The large data sets which we access enable a very fine resolution of heterogeneous agents' behaviour to be included and calibrated in the model. Such an agent-based framework allows us to investigate in detail the effects of multi-agent micro-economic interactions resulting in emergent macro-economic dynamics (*37*). These dynamics are not, in general, susceptible to solutions based on simple analytic techniques. Such canonical methods typically assume normal distributions around equilibrium outcomes as well as linearity of cause and effect (*38*). Agent-based models capture non-linear interactions more naturally by exploiting the fine-grained behaviour calibrated by large data sets.

Agent-based models have already found considerable success in epidemiology (*24*, *39*), social sciences (*22*, *40*), and ecology (*41*, *42*). In this paper we provide further evidence for the applicability of ABMs to the discovery of the mechanisms generating non-linear behaviour in economics.

**Materials and Methods**

*Simulation details*

Each simulation produces a particular trajectory of the price dynamics, which corresponds to the imposed market structure. An ensemble of 1000 trajectories is analyzed for each period to obtain a representative price evolution. Each simulated household represents 10 real households, resulting in more than 200 000 households in the system. The model runs in time steps, which are equivalent to 1 month of real time each. Each simulation consist of the equilibration period and the calendar period. The equilibration period is needed to accommodate biases of initial distribution of wealth and houses between the households. The calendar period follows the real time with a specific starting date. In our model we use the equilibration period of 26 months and the calendar period of 30 months.

In the model we calculate the average monthly housing price of the simulated market. We next compute the 12-steps moving average price, which is referred as the yearly moving average. The range of 12 steps (corresponding to 12 months) is chosen to discard the effect of seasonal price variations.

To calibrate the trend-following aptitude we run simulations for different values of $h$. The median price of the ensemble of 64 trajectories is compared to the observable price dynamics given by the *SIRCA* data on behalf of *CoreLogic Inc*. The resulting value of $h$ is chosen as the one which produced the best fit trajectory by the least squares method. The trajectories for different $h$ and their distance to the CoreLogic trajectory are shown in Fig. S7.

To investigate high price variability in the period of 2016-2019 we have simulated a number of "alternative histories", as defined by alternative sets of input data. Each alternative history is described by almost exactly the same data as the real data of the 2016-2019 period, except the values of one parameter. The alternative values for each of these parameters are taken from the 2011-2014 input data. Plugging the values of each of the alternative parameter into the 2016-2019 data we were able to isolate their effect on the price dynamics, in particular, on its ensemble variability.

*Data description*

The model uses a number of data sources to initialize the agents' behavior, which are listed in Table 2. The data are available in an aggregate format, in term of distributions or time-series, which are illustrated in Supplementary Information. The individual properties of the agents are sampled from the corresponding distributions.

The model uses a number of internal parameters, which reflect characteristics of an individual agent and of the algorithm and are listed in Table 3. The values of these are the same for all periods and therefore reflect the algorithm rather than the actual state of the market. Yet, just like the algorithm is designed to mimic the typical households' behavior, the values of the internal parameters are chosen to account for the typical households' decisions.

The model is calibrated against aggregate housing transactions data, which were synthesized from anonymized individual property transaction records, approved and supplied by *Securities Industry Research Centre of Asia-Pacific* (SIRCA) on behalf of *CoreLogic Inc.*

*The housing market algorithm*

Our ABM for housing market of Greater Sydney is defined by the household agents' decision-making algorithm coupled with the values of the internal parameters representing typical household decisions. The model is coupled with the environmental parameters, which represent the input from the real market. Changing the input allows us to simulate possible effects of various financial and economic policies over time.

The decision model of a household agent is composed of a deterministic component and a stochastic component. The deterministic component is reflected in the behavioral algorithm of an agent and in the values of environmental constraints. The stochastic component is reflected in the heterogeneity of agents' decisions and attributes, which is accounted for by sampling from the specific distributions which represent the actual Sydney housing market and typical household behavior. The result of a single simulation is a particular evolution trajectory of the simulated world, which may be compared to the real world evolution. Yet, because of the presence of a stochastic component, we analyze the evolution of an ensemble of trajectories, which is a collection of the simulation outputs with the same values of the deterministic component.

Every *residential* agent receives monthly *income*, which is accumulated as *wealth*. Furthermore, it may possess a number of *houses*. One of these houses (if there are any) is the agent's *residence*, the other (if there are any) are the *valence* houses, used for investment. In particular, the valence houses are available to other agents without own houses for renting as residences. Renting continues indefinitely until the resident comes in possession of an own house. A house is *owned* by a single owner, either an agent or the *developer* who creates new houses. Every house possesses a single attribute called *quality*, which is defined as the price of the house before the start of the simulation and is kept constant during the simulation. A house may be transferred from the one agent (*seller*) to the other agent (*buyer*), which is compensated by the *price* being transferred from the buyer to the seller. This constitutes a *deal* which is a result of the *market*, and the deal price contributes to the output of the model. Every agent aims to come in possession of at least one house (to be used as a residence) and to maintain a reasonable balance of its own budget, which essentially means attempting to buy a house whenever circumstances allow doing that.

Besides the developer, the model contains three other auxiliary agents, the *overseas* agent, the *bank*, and the *government*. The overseas agent is endowed with a certain *capacity* to buy and rent out valence houses, which is approved by the regulator, and aims to fill that capacity whenever circumstances allow that, acting essentially as a buy-to-let investor. The bank agent can issue an unconstrained amount of lending to residential agents in the form of the *mortgages*. The government applies taxes and determining the capacity of the overseas agent.

At every month $t$ the residential agents update their income $I$ and wealth $W$ according to Eq. (1), which sets the income growth and the budget accounting rules. Next, the agents participate in the market, which consists of several steps. First, every residential agent chooses the minimum value of $P_1$, $P_2$, $P_3$ prices defined by Eq. (3) as the *bid* price $P_b$, and, if $P_b/P_1$ is larger than the *expectation downshift* (willingness to downgrade the pre-purchase expectation in house price/quality under financial prudence measures), puts the bidding record on the market. This process captures the analysis of the agent's affordability and negotiations with the bank. The overseas agent simply puts the bidding records according to its capacity. Next, all agents go through their owned houses and with probability of 1 % put the listing record on the market with the *list* price defined by Eq. (4). The market is cleared within one-step process with the highest price preference. The listing agents act in the order of the list price to find the corresponding bidding agent with the bid price higher than the list price. If a match is found, then the deal is cleared with the probability of 80 %, which changes the ownership of the house and financial states of the buyer and seller. If no match is found, the listing record is transferred to the next month.

In the following equations we use a simplified notation for the ease of reading. The environmental parameters are denoted by the Greek letter $\Phi$ with corresponding subscripts and are listed in Table 1. They are observed in reality, are the same for all agents but are different across the simulation periods. The internal parameters are denoted by the Latin letter $b$ with corresponding subscripts and listed in Table 3. They are the same across all simulated periods. The internal parameters also account for heterogeneity of the agents, which means that the actual value for a particular agent is sampled from a uniform distribution with a fixed mean and width. The other system variables are denoted by Latin capital letters, e.g. $Q$, $M$, etc. If a variable depends on time within a particular period, this is indicated explicitly by the argument $[t]$.

The income $I[t]$ and the wealth $W[t]$ are updated according to the following rules:

(1)
$$I[t] = (1 + b_I)I[t-1]$$

$$W[t] = (1 - b_{CW})W[t-1] + (1 - b_{CI})(1 - \Phi_T)I[t] - R_r - \sum_{h=1}^{H}(\Phi_H Q_h + M_h[t] - R_h)$$

Here $H$ is the number of the owned houses, while $h$ is the house number, $Q_h$ is the quality of the house and $M_h$ is the mortgage payment for the house. Furthermore, $R_i$ is the rental payment (if applicable), which is subtracted from the wealth if the house is the residence $r$ and is added to the wealth if the house is an owned valence house $h$. Rent is calculated as

(2)
$$R_h = \frac{1}{3}\Phi_R + \frac{1}{3}b_{RI} I + \frac{1}{3} b_{RH} M_h$$

where $\Phi_R$ is a random value drawn from the rent bracket distribution, sourced from Census for each simulated period.

The bid price is chosen as the minimum of three alternative candidates:

(3)
$$P_1[t] = \frac{b_b \Phi_b (I[t])^{\Phi_I} U_b[t]}{\Phi_{LTV} \Phi_M[t] + \Phi_H - h \Delta_{HPI}[t]}$$

$$P_2[t] = \frac{b_{ATW} W[t]}{1 - b_{LTV}}$$

$$P_3[t] = \frac{b_{DTI} I[t]}{b_{LTV} b_M}$$

where $U_b$ is the urgency of the agent to buy a house, which is different from 1 when the agent has recently sold a house and has excess cash.

The list price is determined as:
(4)

$$P_\ell[t] = \frac{b_\ell \overline{Q_h} \, (S[t])^{b_s} \, U_\ell[t]}{(1 + D_h[t])^{b_d}}$$

where $b_\ell$ is the listing premium. Furthermore $U_\ell$ is the urgency of the agent to sell a house, which is different from 1 either when the agent is financially stressed or when his valence house is not rented out. $S[t]$ is the market average of the sold-to-list price ratio, while $D_h$ is the amount of months the house has been listed on the market. Finally, $\overline{Q_h}$ is the average quality of the 10 most similar houses.

**Acknowledgments**: Authors thank Markus Brede and Doyne J. Farmer for useful comments on the paper draft. Data on housing transactions are supplied by Securities Industry Research Centre of Asia-Pacific (SIRCA) on behalf of *CoreLogic, Inc.* (Sydney, Australia). We also acknowledge the Australian Bureau of Statistics for the access to Census data and the Melbourne Institute for the access to the "Household, Income and Labour Dynamics in Australia" Survey. **Funding:** This work is supported by the Australian Research Council Discovery Project DP170102927 for MH and MP and by University of Sydney Mobility Scheme-2019 for KG. **Author contributions:** KG analyzed the source data, developed the software code, performed and analyzed the simulations, prepared the manuscript ("Results" and "Materials and Methods" sections, Figures and Tables); KG, MP and MH developed the model; AC consulted on agent-based modelling; PO consulted on economic modelling; all authors contributed to "Introduction" and "Discussion" sections of the manuscript. **Competing interests:** Authors declare no competing interests. **Data and materials availability:** All data needed to evaluate the conclusions in the paper are present or referred to in the paper. The reference data for housing transactions are owned by *CoreLogic, Inc.*. Additional information related to this paper may be requested from the authors.

**Figures and Tables**

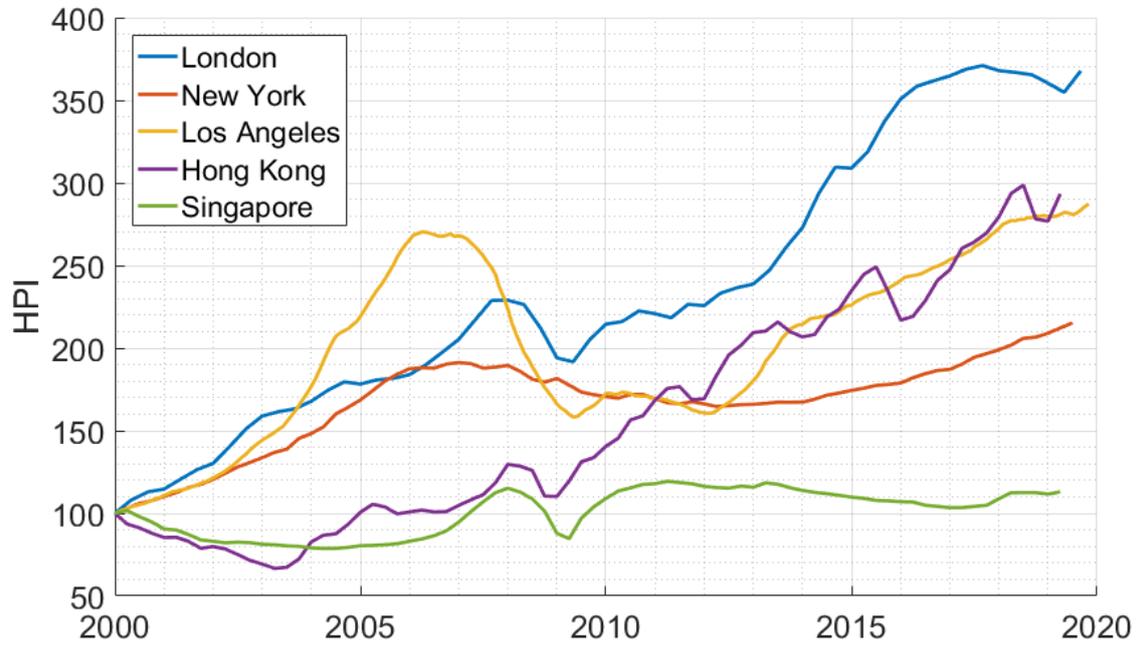

**Fig. 1. House price index.** The quarterly house price index for selected cities across the world during the past 20 years, normalized at 100 in year 2000. Sources: *https://fred.stlouisfed.org* and *https://www.gov.uk*.

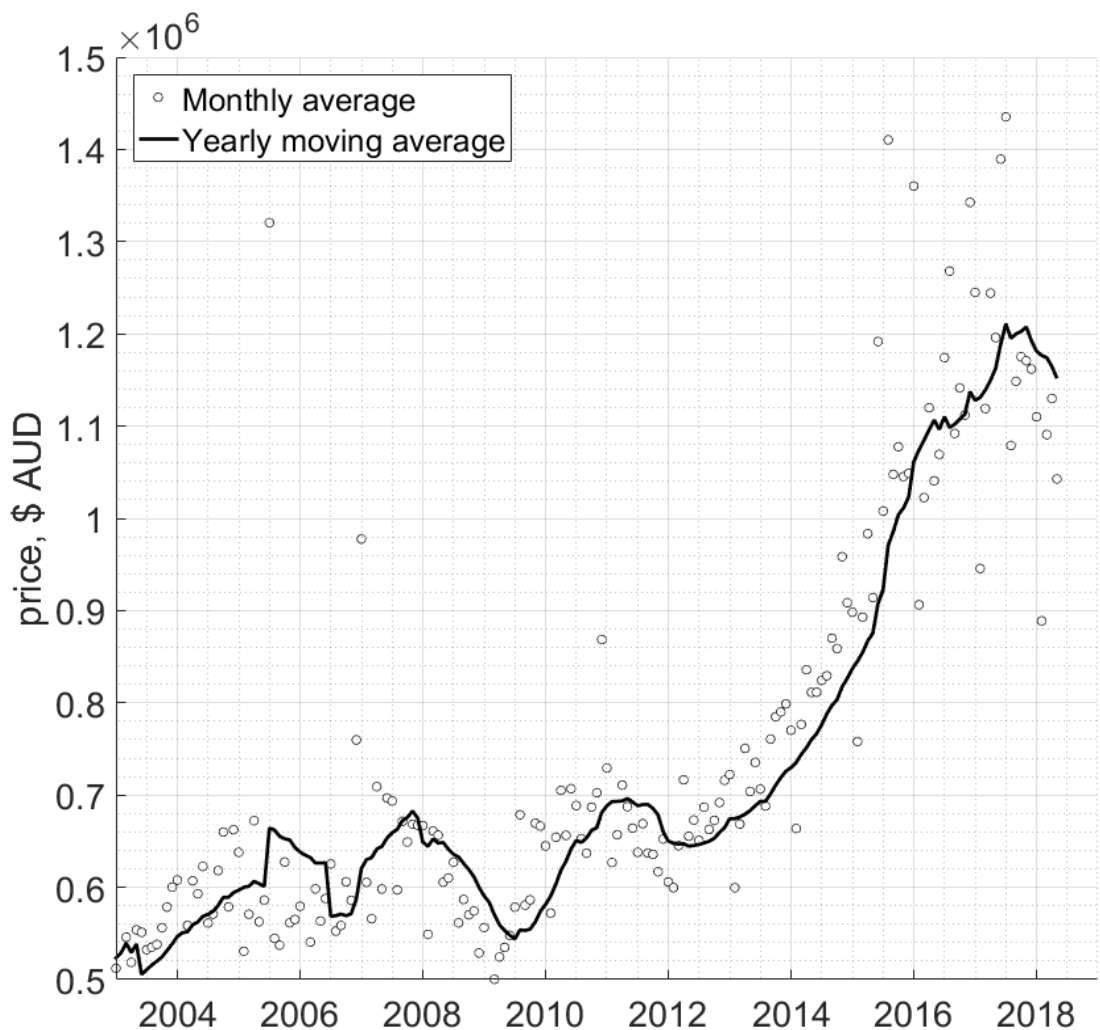

**Fig. 2. Greater Sydney house price.** The monthly average price (circles) and the yearly moving average price (line) of the housing sales for Greater Sydney region. Source: *Securities Industry Research Centre of Asia-Pacific* on behalf of *CoreLogic, Inc.*

**Fig. 3. Agent-based model output.** The market price produced by multiple simulations, for the period of 2006-2009 (top), the period of 2006-2009 (middle), the period of 2006-2009 (bottom). For each period, the left figure shows 1000 simulated trajectories (thin solid colored lines), together with the median trajectory (thick yellow line) and the monthly averages of the actual sale prices (CoreLogic: black circles with a thin dashed line). The right figure shows the histogram distribution of the ensemble (color shade), together with the yearly moving average of sale prices (CoreLogic: dashed black line).

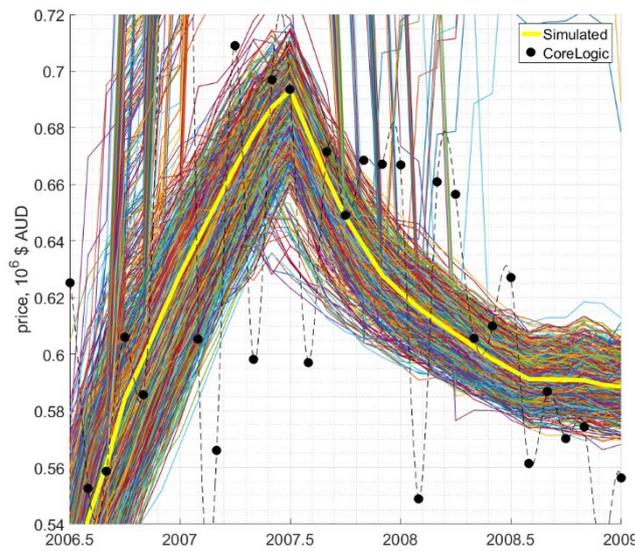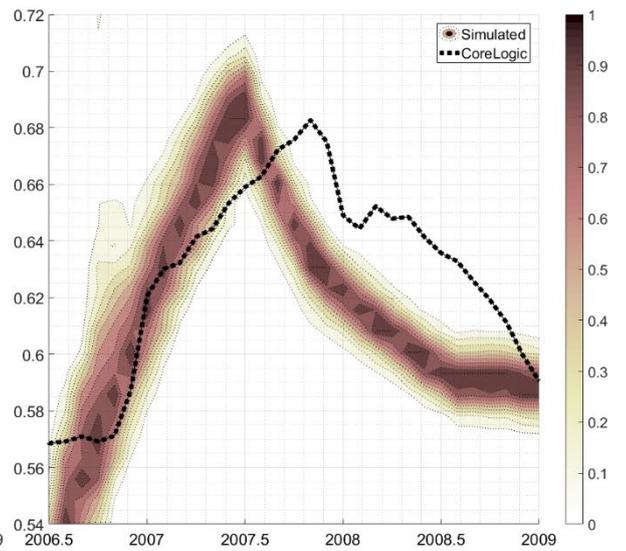

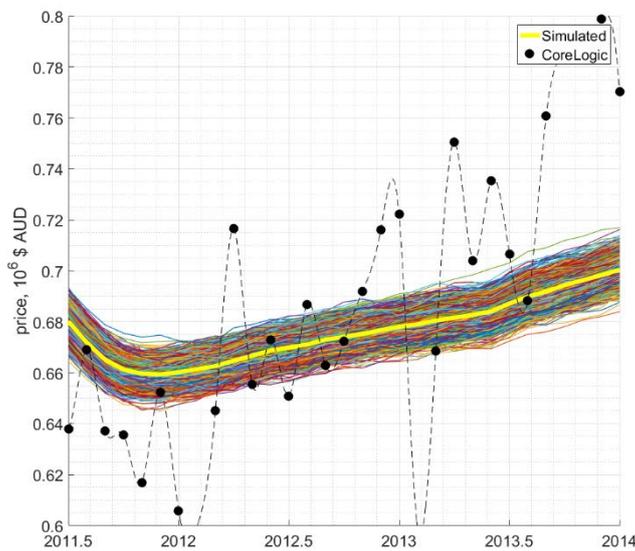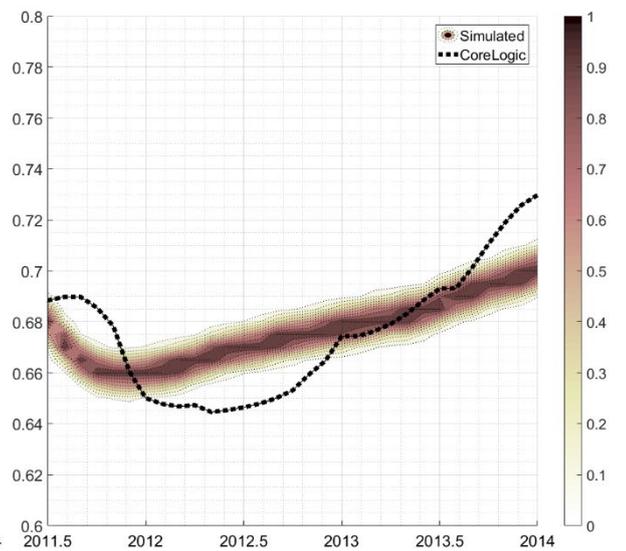

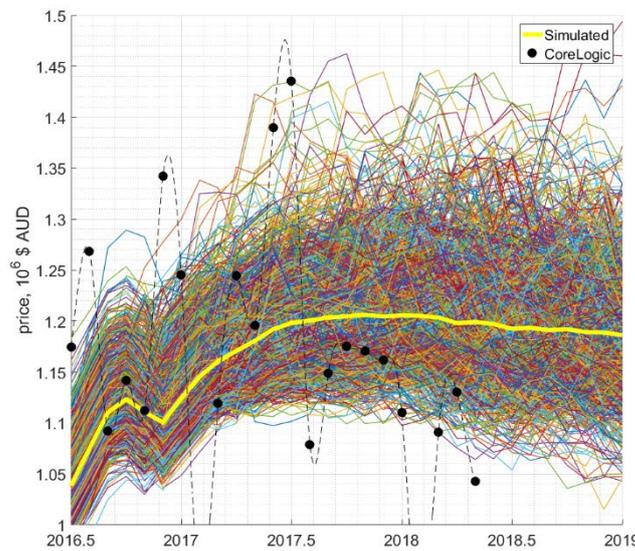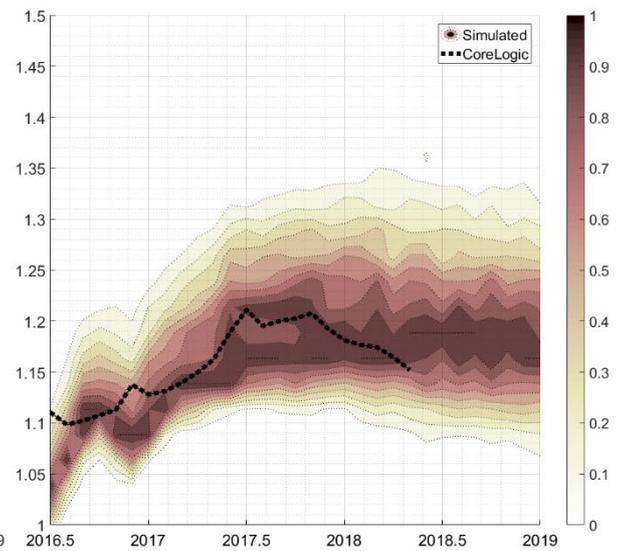

| Name | Symbol | Values | | |
|---|---|---|---|---|
| | | 2006-2009 | 2011-2014 | 2016-2019 |
| Income tax brackets, $'000 | | 6, 25, 75, 150 | 6, 37, 80, 180 | 18.2, 37, 87, 180 |
| Income tax rates, % | $\Phi_T$ | 15, 30, 40, 45 | 15, 30, 37, 45 | 19, 32.5, 37, 45 |
| House owning expenses, % | $\Phi_H$ | 4.2 | 4.2 | 4.2 |
| House purchase tax, % | - | 5 | 5 | 5 |
| Annual mortgage rate, % | $\Phi_M$ | 7.3 – 9.45 | 5.53 – 7.79 | 4.95 – 5.35 |
| Mortgage duration, years | - | 30 | 30 | 30 |
| Mortgage LVR mean, % | $\Phi_{LTV}$ | 72.5 | 67.5 | 60 |
| Mortgage LVR variance, % | - | 12.5 | 12.5 | 12.5 |
| Mortgage-Income statistics | $\Phi_b$; $\Phi_I$ | 689.53; 0.81 | 1072.1; 0.75 | 1141.7; 0.80 |

**Table 1. External parameters of ABM.** The parameters describing external economic and financial conditions of each period used in simulations. These percentages are converted to fractions when used in equations.

| Name | Symbol | Source |
|---|---|---|
| Demographics time-series | - | ABS: tables 3301.0, 3302.0, 3412.0 |
| House construction time-series | - | ABS: tables 8731.0 |
| Foreign investment capacity | - | FIRB: 2011-2016 annual reports |
| Wealth bracket distribution | $W$ | HILDA |
| Income bracket distribution | $I$ | ABS: Census; tables 6523.0 |
| Rent bracket distribution | $\Phi_R$ | ABS: Census |
| Mortgage bracket distribution | $M_h$ | ABS: Census |
| Housing transactions | - | SIRCA / CoreLogic |

**Table 2. Data sources.** Temporal and statistical data used in the model together with their sources. Sources: *Australian Bureau of Statistics* (ABS), *Australian Foreign Investment Review Board* (FIRB), *The Household, Income and Labour Dynamics in Australia Survey* (HILDA), *Securities Industry Research Centre of Asia-Pacific* (SIRCA) on behalf of *CoreLogic, Inc.*

| Name | Symbol | Mean | Half-width |
|---|---|---|---|
| Monthly income growth rate | $b_I$ | 0.002 | 0.001 |
| Fraction of income to pay non-housing consumption | $b_{CI}$ | 0.6 | 0 |
| Fraction of wealth to pay non-housing consumption | $b_{CW}$ | 0.0025 | 0 |
| Fraction of income to spend on rental | $b_{RI}$ | 0.2 | 0.1 |
| Fraction of mortgage payments to be compensated by rent | $b_{RH}$ | 1.25 | 0.25 |
| Bank downpaymet-to-wealth ratio threshold | $b_{ATW}$ | 0.9 | 0 |
| Bank loan-to-value ratio threshold | $b_{LTV}$ | 0.9 | 0 |
| Bank debt-to-income ratio threshold | $b_{DTI}$ | 0.4 | 0 |
| Approval mortgage rate | $b_M$ | 0.07 | 0 |
| Expectation downshift | - | 0.6 | 0 |
| Bid price factor | $b_b$ | 1.29 | 0.29 |
| List probability (fraction) | - | 0.01 | 0 |
| List price factor | $b_\ell$ | 1.70 | 0.5 |
| Sold-to-list price ratio exponent | $b_s$ | 0.22 | 0 |
| Months listed on market exponent | $b_d$ | 0.01 | 0 |
| Clearance probability (fraction) | - | 0.8 | 0 |

**Table 3. Internal parameters of ABM.** Parameters specifying agents' behavior, with their median values and the half-width of the interval, used in the uniform sampling of heterogeneous properties of the household agents.

**Supplementary Materials**

**Input Data**

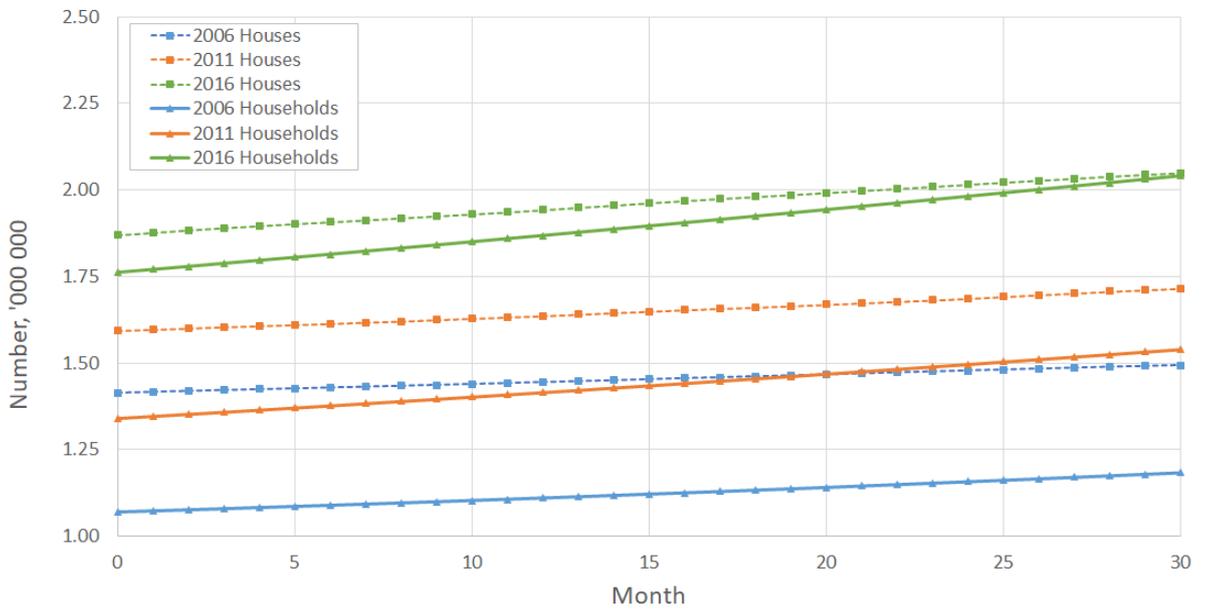

**Fig. S1. Demographics.** Historical number of households (interpolated) and houses. Source: *Australian Bureau of Statistics* (Table 2).

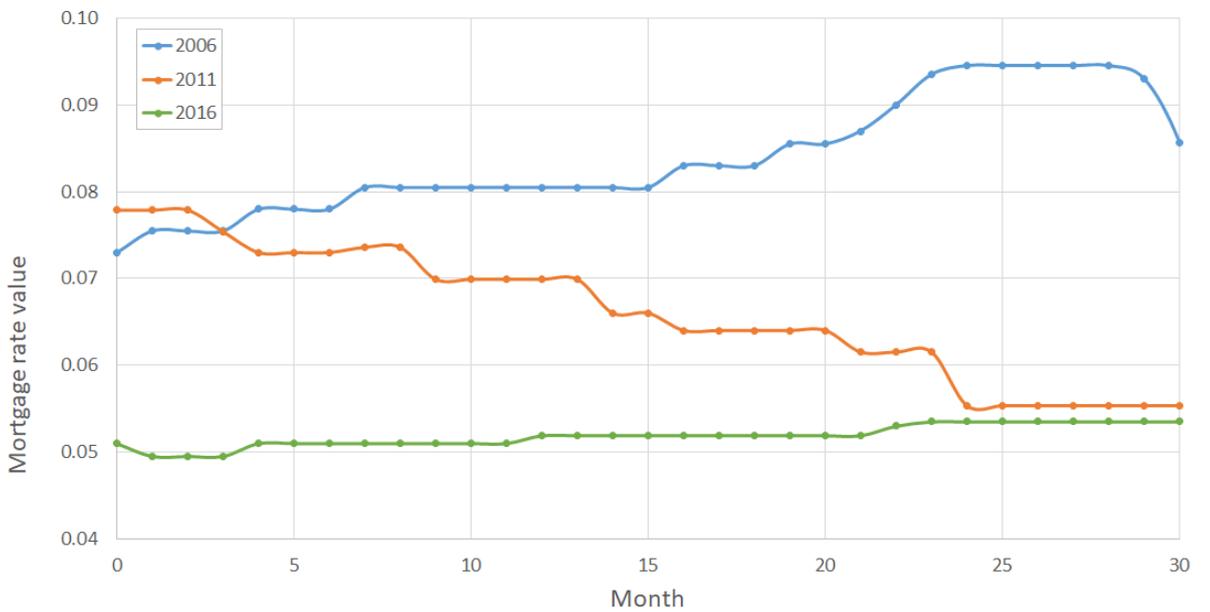

**Fig. S2. Mortgage rate.** Historical mortgage rate values. Source: *https://www.loansense.com.au*.

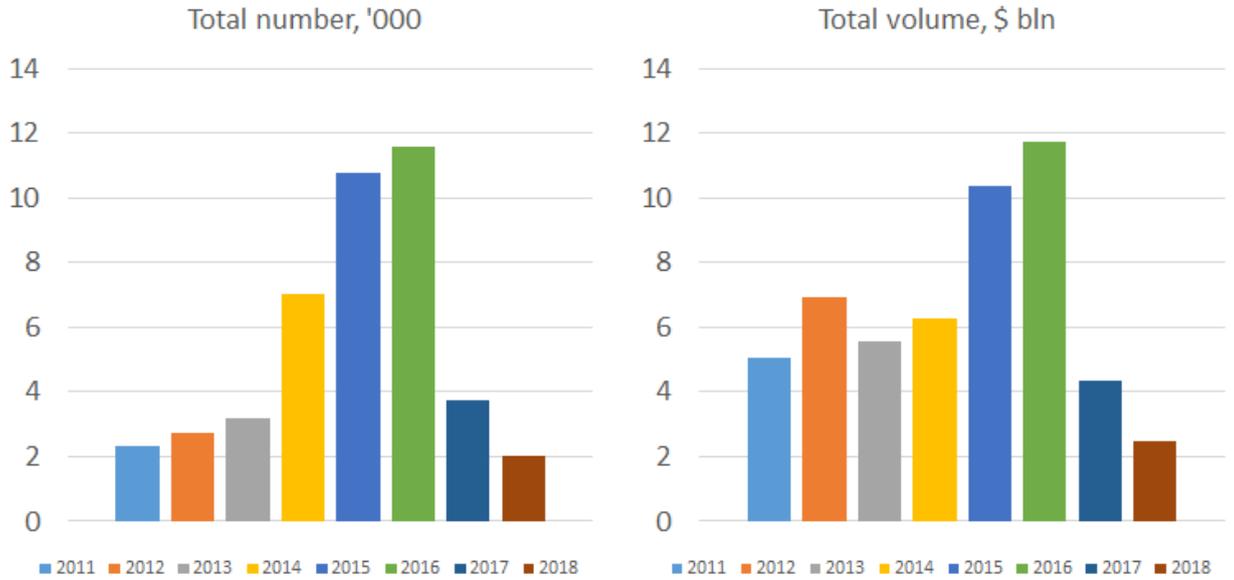

**Fig. S3. Overseas investment.** Historical capacity of the total number of sales and volume of the overseas buying in New South Wales. Source: *FIRB* (Table 2 in main text).

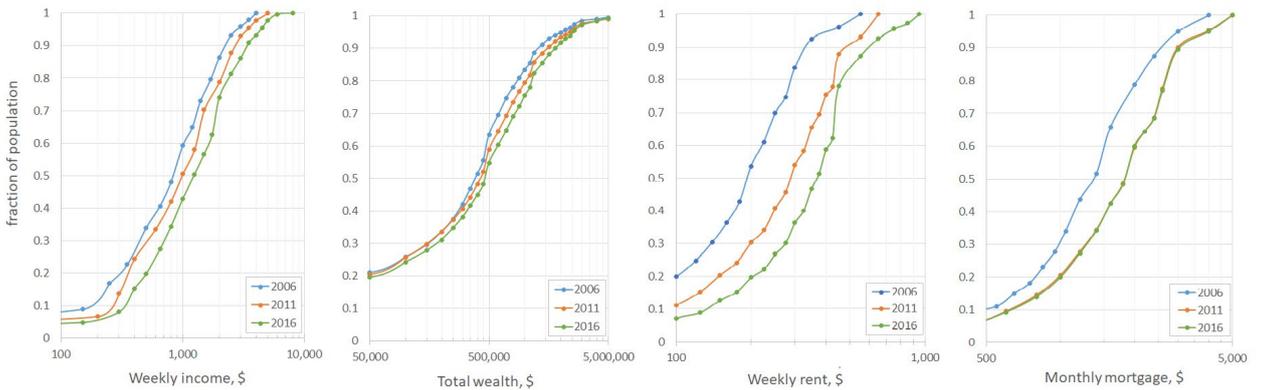

**Fig. S4. Household financial distributions.** Distributions of income, wealth, rental and mortgage payments. Source: *Australian Census* (income, mortgage, rental), *HILDA* (wealth), see Table 2 in main text. The points on the curves indicate the brackets' lower bound of the corresponding bracket distribution.

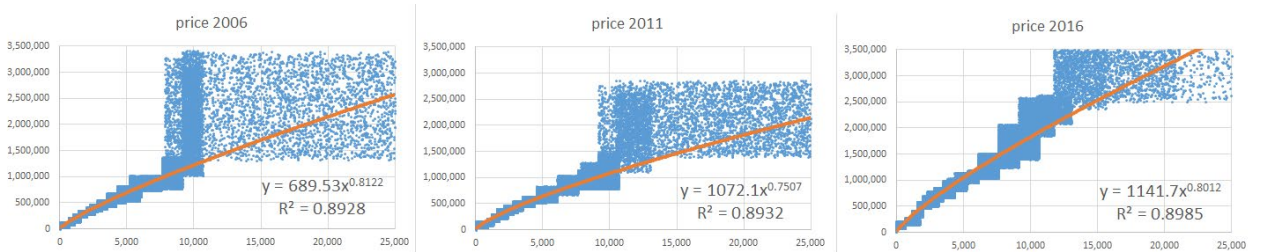

**Fig. S5. Mortgage-Income distribution.** Regressions generated from Mortgage and Income distributions (Table 2 in main text) to evaluate $\Phi_b$ and $\Phi_I$ (Table 1 in main text) for 2006-2009 (left), 2011-2014 (middle), 2016-2019 (right).

# Validation Analysis

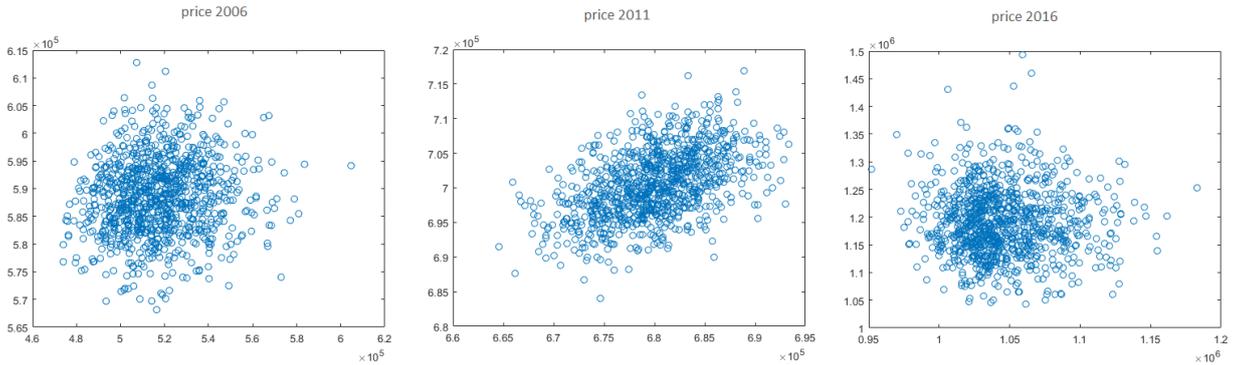

**Fig. S6. Validation of ensemble independence.** The price in the end of the simulated period (vertical axis) against the price in the beginning of the simulated period (horizontal axis), for 2006-2009 (left), 2011-2014 (middle), 2016-2019 (right).

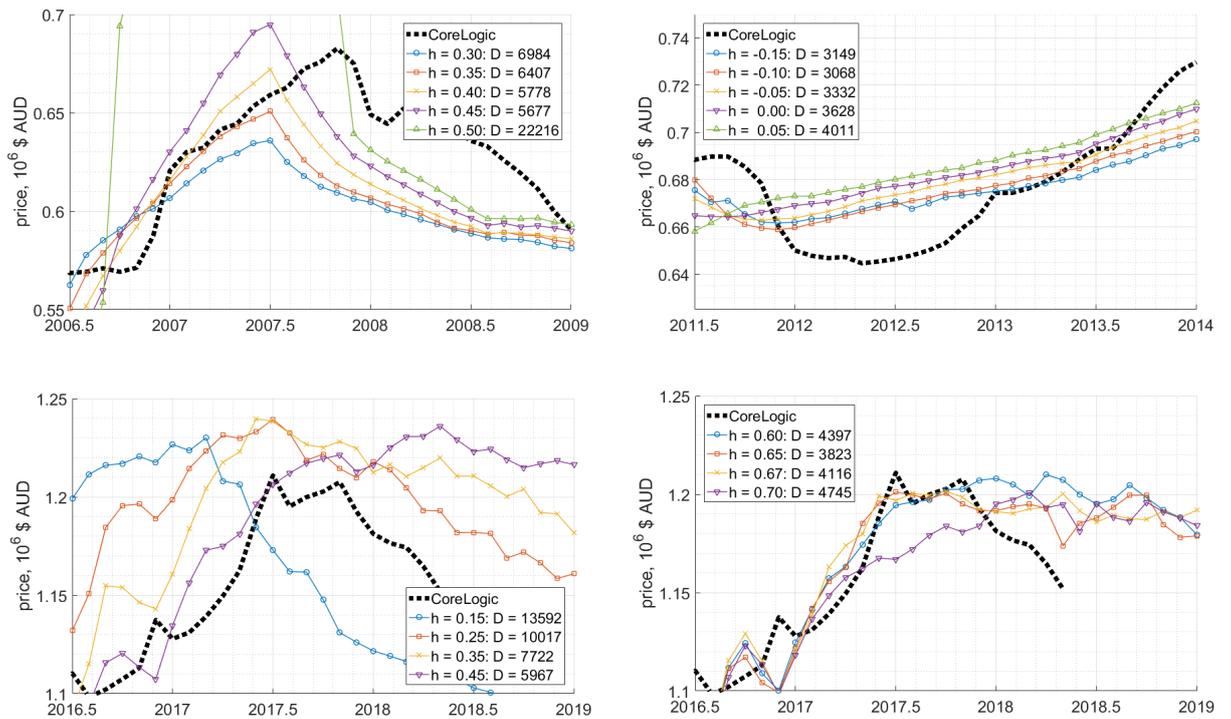

**Fig. S7. Calibration of the trend-following aptitude.** The average price of the simulated ensemble of 64 trajectories for different values of the trend-following aptitude $h$, together with the mean-squared distance $D$ to the actual price trajectory, for 2006-2009 (top left), 2011-2014 (top right), 2016-2019 (bottom, right and left). The calibrated value of $h$ is the one for which $D$ is the smallest.

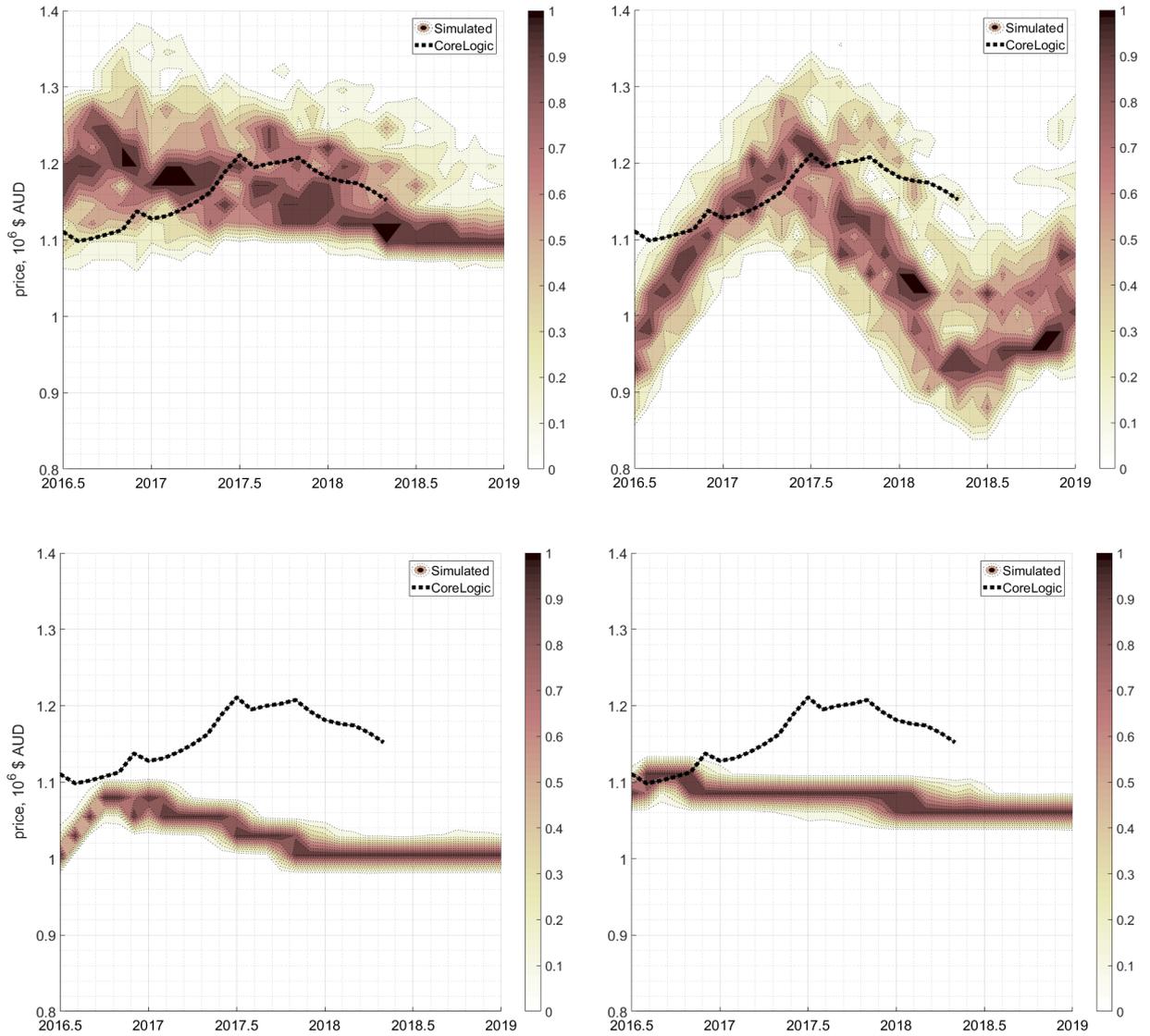

**Fig. S8. Testing alternative hypotheses for 2016 price.** The model output for the "alternative history" period of 2016-2019, for which $h = 0.20$ is set instead of $h = 0.65$, while some of the input data are taken from the 2011-2014 period. Top left: baseline simulations with all of the input from the 2016-2019 period. Top right: simulations with the initial price level from the 2011-2014 period. Bottom left: simulations with the mortgage-income distribution from the 2011-2014 period. Bottom right: simulations with the mortgage rate from the 2011-2014 period.